\def\cm2{cm$^{-2}$}

\def\c2{C~{\sc ii}}
\def\c4{C~{\sc iv}}
\def\fe2{Fe~{\sc ii}}
\def\fe3{Fe~{\sc iii}}
\def\mg1{Mg~{\sc i}}
\def\mg2{Mg~{\sc ii}}
\def\si2{Si~{\sc ii}}
\def\si4{Si~{\sc iv}}
\def\al2{Al~{\sc ii}}
\def\al3{Al~{\sc iii}}
\def\o1{O~{\sc i}}
\def\n1{N~{\sc i}}
\def\h1{H~{\sc i}}

\def\approxlt{\mathrel{\spose{\lower 3pt\hbox{$\sim$}}
        \raise 2.0pt\hbox{$<$}}}
\def\approxgt{\mathrel{\spose{\lower 3pt\hbox{$\sim$}}
        \raise 2.0pt\hbox{$>$}}}

\documentclass[article]{gwp80}
\usepackage{graphicx}
\usepackage{epstopdf}
\usepackage{amssymb}
\tabletypesize{\normalsize}  

\shortauthors{Rich}
\shorttitle{The BRAVA Survey}

\begin{document}
\large    
\pagenumbering{arabic}
\setcounter{page}{264}

\title{The Bulge Radial Velocity/Abundance Assay}

%
%
\author{{\noindent R. Michael Rich {$^{\rm 1}$}\\
\\
{\it (1) Department of Physics and Astronomy, University of California, Los Angeles}
}
}
%
%
\email{(1)rmr@astro.ucla.edu}

\begin{abstract}
The Bulge Radial Velocity/Abundance Assay (BRAVA) has accomplished a survey of 10,000
 red giants in the Southern Galactic bulge, approximately spanning $ -8^\circ < l < +8^\circ$
 and $-3^\circ <b < -8^\circ$, a region within roughly 1 kpc from the nucleus.  We
 find that the Galactic bulge at $b=-4^\circ$ displays a clear departure from solid
 body rotation, and that the rotation field along the major axis at $b=-6^\circ$ and
 $b=-8^\circ$ is identical to that at lower latitude; this is ``cylindrical'' rotation,
 a hallmark observed in edge-on bars.  Comparison of the BRAVA dataset with
 an N-body bar shows that $>90\%$ of the bulge population is in the bar, leaving little
 room for a ``classical'' bulge component.  We also report on the first iron abundance
 and composition measurements in the outer bulge, at $b=-8^\circ$.  The iron abundance
 in this field falls on the trend of a suspected gradient measured from high resolution
 spectroscopy of bulge clump stars.  Further, we find that the trends of $\rm [\alpha/Fe]$
 vs [Fe/H] that characterize the bulge at lower latitude are present 1 kpc from the
 nucleus, consistent with a rapid $(<1 \rm Gyr)$ timescale for the formation of the
 bulge, even near its boundary.   Although the dynamics of the bulge are consistent
 with those of a dynamically buckled N-body bar, the presence of an abundance gradient
 is not compatible with purely dynamical processes; we propose that missing baryonic
 physics is needed.  We also report on the remarkable massive bulge globular cluster
 Terzan 5, which has a bimodal abundance and composition distribution, and is proposed
 as the remnant of a population of primordial building block stellar systems that formed
 the bulge.   Terzan 5 is presently a unique case, and it is important
 to test whether the dissolution of systems similar to it populated the bulge.

\end{abstract}

\section{Forward}

My connection with George Preston extends back to my doctoral student days at Caltech.
 My doctoral thesis at Caltech (with Jeremy Mould as advisor) was on the abundances and kinematics
 of K giants in the Galactic bulge, and I took data on the then newly commissioned
 100-inch du Pont telescope at Las Campanas.   The Carnegie Director was George Preston;
 I enjoyed substantial allocations of observing time, although much time was lost to
 the clouds and winds of June and July, typical for the Chilean winter.   Without his
 support and that of the outstanding personnel at Las Campanas, and without Steve Shectman's
 Shectograph instrument, my career in astronomy would not have been possible.  The
 most distinct difference between the Las Campanas of ``then'' and now is that communication
 with the outside world was by shortwave radio, only.  It was possible to phone to
 the US only be calling by radio the Pasadena headquarters, where a ``phone patch''
 connected one to the US mainland.   There was, of course, no internet connection.  Cloudy nights
 were spent either reading the literature, or browsing a two foot deep pile of New
 Yorker magazines in the 100-inch library.  Las Campanas was truly a world unto itself,
 with its own generators, no phone lines, and link to the world via carryall and shortwave
 radio.  

My other recollection was the warmth and self-effacing character of George Preston,
 which contrasted so strongly with other senior astronomers of the time.  While maintaining
 the highest standards of scientific and administrative excellence, he was able to
 laugh at himself and never took himself too seriously.   George could turn any situation into a good
 chuckle.  As a Southern California native, I
 did not know much about driving in snow.  George watched me attempting to back out
 of a parking space and navigate the (then) newly paved roads of Las Campanas-all this,
 after one of the winter snows that sometimes hit Chile that time of year.  I finally
 managed to complete the vehicular maneuvers, while George enjoyed a good belly laugh.
  There is a serious side to all of this, though.  George's model of scientific achievement
 and genuine discovery at epochal levels, combined with modest, warmth, and accessibility
 to young people, is one our field of today could sorely use.    
 
 \subsection {Introduction}

The dominance of M giants in the Galactic bulge has been one of the longest established
 facts regarding that stellar population; it was first noted in the objective prism
 surveys of Nassau \& Blanco (1958).  When Blanco moved to Cerro Tololo, among the
 first projects to be undertaken was use of the prime focus grism at the 4m
telescope
 to obtain low dispersion classification spectra of fields in the Magellanic Clouds
 and the Galactic bulge; it was almost immediately discovered that the bulge was completely
 dominated by M giants and lacked carbon stars, in contrast to the Magellanic Clouds.
  Victor and Betty Blanco made a critical contribution that fueled an important surge
 in the understanding of the AGB phase of evolution, and it is one that continues to
 have impact today.  Blanco, McCarthy, \& Blanco (1984) obtained spectral types for
 M giants in the bulge, {\it and} precise coordinates.  They did the same for the Magellanic
 Clouds as well.  It was this work that enabled the landmark paper of Frogel \& Whitford
 (1987) on the infrared properties of the M giant population (and leading to the bolometric
 magnitudes and effective temperatures of the stars, relative to the globular cluster
 population).  It is noteworthy that the deep understanding of the AGB phase, along
 with important theoretical work (e.g. Iben \& Renzini 1983) began a trajectory of
 research that has impact on the understanding of high redshift galaxies.  It will
 also come to the fore when the James Webb Space Telescope undertakes studies of nearby
 stellar populations.  Thanks to the ``low technology'' of the spectral classification
 from the grism spectra, and hand-measured astrometric positions, the raw material
 for this surge in knowledge was in place.  Suddenly, one had in place coordinates
 for late type stars in globular clusters, the bulge (metal rich and old) and the Magellanic
 Clouds (intermediate age).  All of this came online just as infrared measurements
 became widely available, and Aaronson, Mould, and Frogel were able to write their
 series of important papers.

Motivated by the copious numbers of M giants in the bulge, Mould (1983) used the population
 to measure the first velocity dispersion (from Las Campanas, with the du Pont telescope).
  Shortly afterward, Sharples, Walker \& Cropper (1990; SWC90) used the first AAT fiber-fed
 spectrograph to measure 239 radial velocities for Blanco M giants in Baade's Window.
  Had it not been for the grism spectra and the precise coordinates measured from photographic
 plates, this project would have been impossible.  In fact, despite the ease with which
 bulge M giants could be identified and applied as kinematic probes, progress stalled
 for 15 years; no attempt was made to do multi-object spectroscopy on K- or M- giants in the Galactic
 bulge, save for the innovative Fabry-Perot imaging technique using a Ca triplet line (Rangwala \& Williams 2009).

The author conceived the BRAVA project (Rich et al. 2007a; Rich et al. 2009) because
 the 2MASS survey provided an unlimited database of bulge M giants with excellent
 positions, and the stars are surprisingly bright in the I band (the brightness distribution
 of the SWC90 sample peaks at I=13, for Baade's Window).  The TiO bands and Ca infrared
 triplet are superb for cross correlation at a wide range of spectral resolutions.
  Our team employed the CTIO 4m hydra echelle spectrograph in its low resolution mode,
 affording spectroscopy of $\approx 100$ stars at each pointing.  Because of concerns
 about strong night sky lines in the Ca triplet region (with consequent subtraction
 issues using fibers) the initial datasets stopped blueward of the Ca triplet.  However,
 from 2007 onward, all data included at least one of the Ca infrared triplet lines.

The primary goal of BRAVA is to provide a fundamental dataset of M giant kinematics
 that extends over as much of the bulge as possible.  Easily selected from the red
 giant branch (Figure 1) the M giant sample suffers less potential contamination than
 the red clump because of the sloping nature of the red giant branch, that permits
 a well defined selection region.  The final positions of fields with data are shown
 in Figure 2.   Although there have been serious efforts made to compare bulge datasets
 with dynamical models (Beaulieau et al. 2000), the PNe used in that study might suffer
 from disk population contamination, and are in any case few in number (1000 vs the
 10,000 star sample of BRAVA).    

A secondary goal of BRAVA, as mentioned in the title, is a survey of the bulge to characterize
 the element abundances of the stars.  This part of BRAVA is being led by Christian
 Johnson, and will result in new determinations for alpha and heavy element abundances
 along the bulge minor and major axes, and in the thick disk.   At present, observations of
 roughly 500 stars in 3 fields along the major axis have been obtained; the Plaut field has been
 analyzed and is reported in the literature (Johnson et al. 2011).

\begin{figure*}
\centering
\includegraphics[width=0.90\columnwidth]{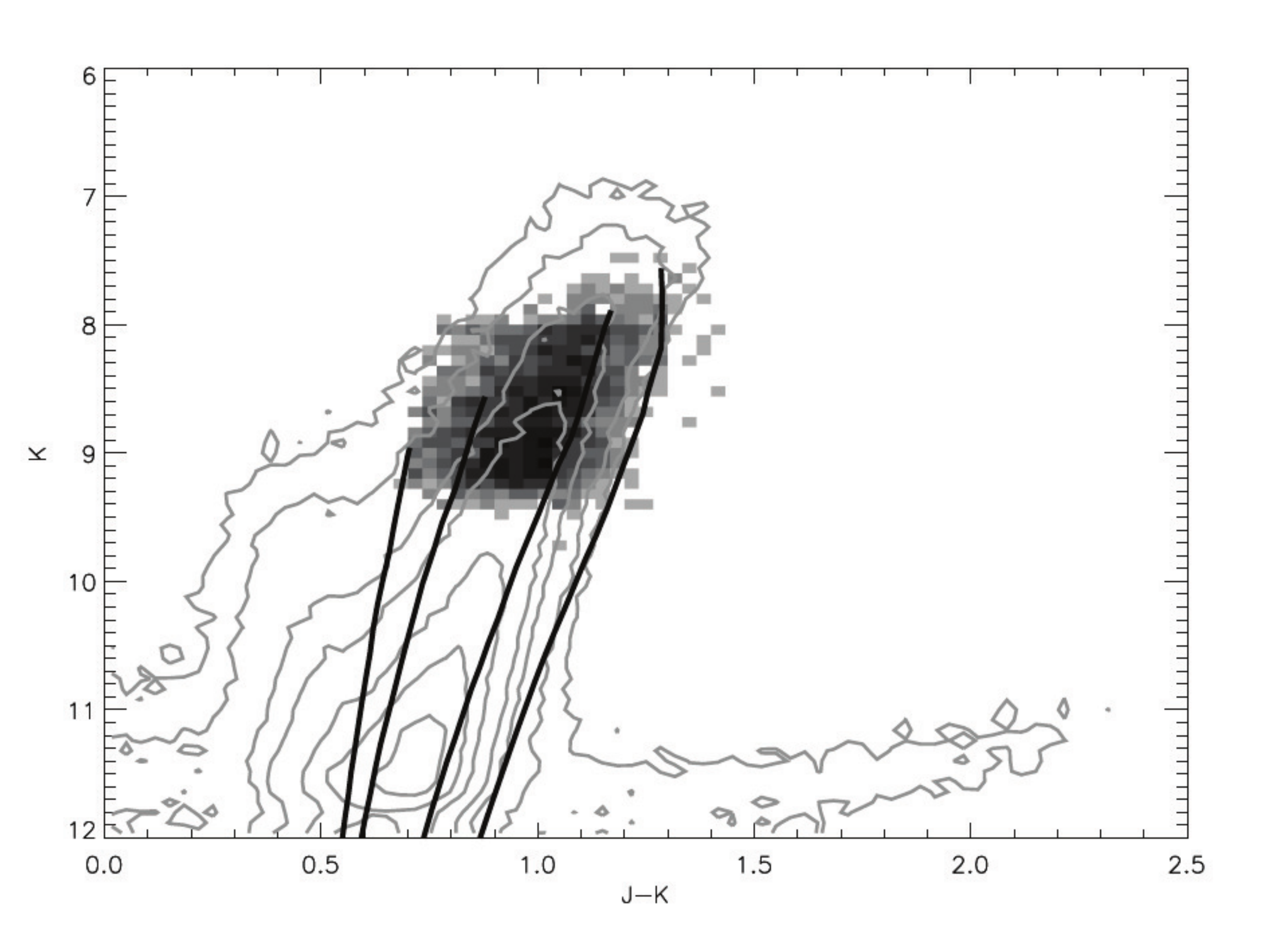}
\vskip0pt
\caption{Color-Magnitude contours of the 2MASS catalog ($\sim$440,000 stars) for 
fields along $b=-4^\circ$ major axis strip (Howard et al. 2008).  Superimposed in greyscale are our
observed BRAVA targets (deredenned),
representing 2505 stars, and isochrones for a 12 Gyr population at a distance modulus of
14.47 mag. The isochrones, starting on left, are for [Fe/H]= $-2.0,-1.3,-0.5$, and $+$0.2
(Marigo et al. 2008).   We have excluded roughly 100 stars bluer than the [Fe/H]=$-2$ isochrone, 
and $K<7.4$. }
\label{BRAVA sample and Globular Cluster Fiducial Giant Branches}
\end{figure*}

\begin{figure*}
\centering
\includegraphics[width=0.90\columnwidth]{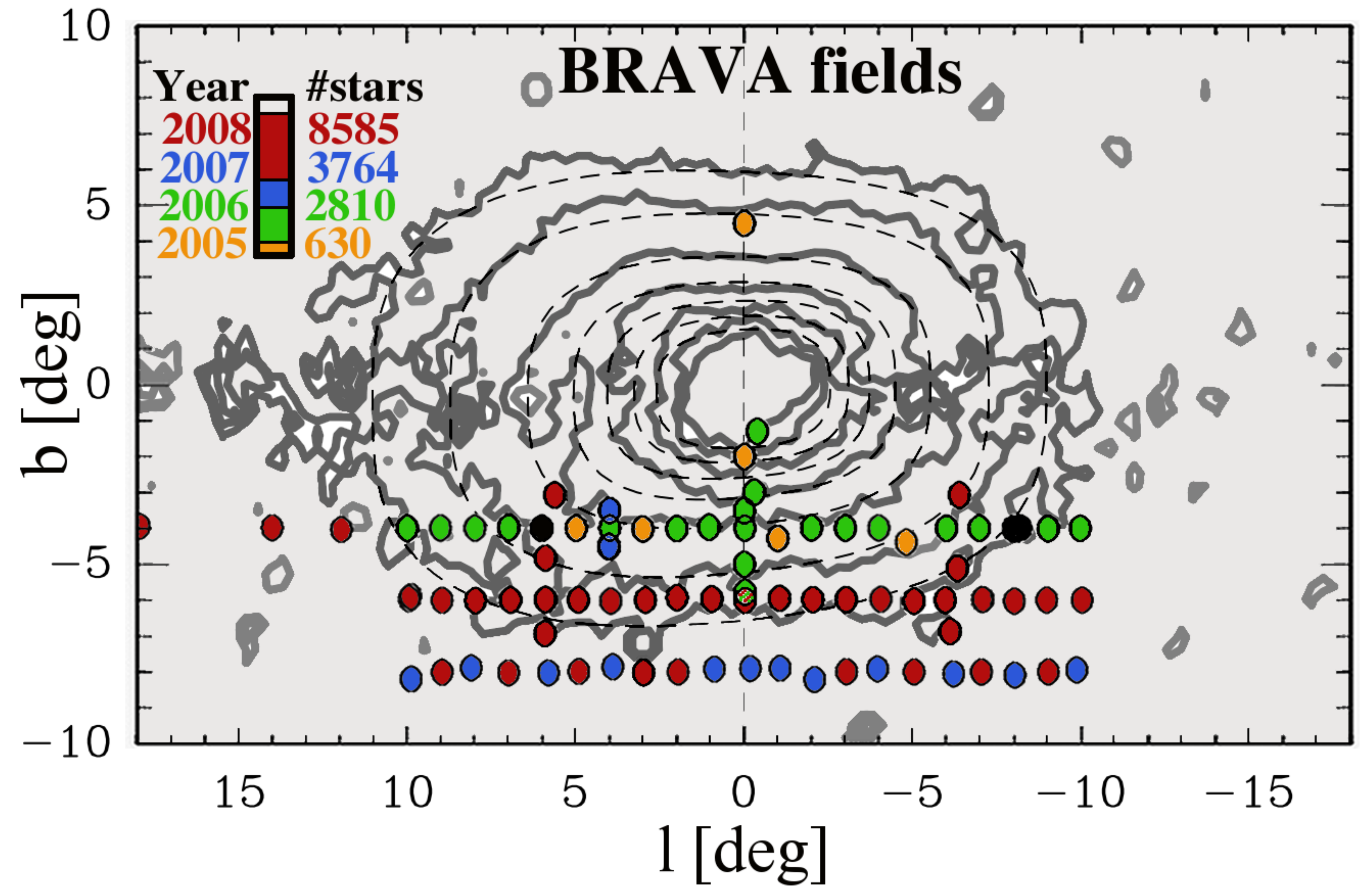}
\vskip0pt
\caption{ Final configuration of fields of observed for the low resolution (R$\sim 4000$) hydra
campaign, courtesy A. Kunder.  It would have been feasible to cover the Northern Galactic bulge and higher latitudes,
but time was not granted to do so.}
\label{BRAVA sample and Globular Cluster Fiducial Giant Branches}
\end{figure*}

\vbox{
\vskip24pt
}
\section{Issues Concerning the Formation of the Bulge}

Broadly speaking, we note a bulge as a central concentration that is accompanied 
by a disk or dust lane.  An elliptical galaxy, or S0, with no hint of dust and star
formation does not properly have a bulge; it is a spheroid.   M104 has a bulge, 
because it also has a disk of dust and star formation.   However, the bulge of M104
so dominates that galaxy and extends into the halo, that one might consider it 
properly to be a bulge/halo with a weak disk, rather than a kind of spiral galaxy.  

Within spiral galaxies, we now distinguish between classical bulges and pseudo-bulges,
a class that also includes bars (Kormendy \& Kennicutt 2004).   Our Milky Way is likely
a galaxy that hosts a pseudo-bulge.

The modern theory of bulge formation is basically divided between two broad ideas.  The
merger-driven early bulge, driven by the baryonic processes associated with the accretion
of major clumps of dark matter and gas (e.g. Abadi et al. 2003; Elmegreen et al. 2008).
This is the a more modern version of the classical Eggen, Lynden-Bell, \& Sandage (1962) model that
involved violent relaxation.   The merger-driven model is understood within the context of
the LCDM galaxy formation picture.

The second broad class of models is those involving secular evolution (Combes \&Sanders 1981; Raha et al. 1991; Norman et al. 1996; Athanassoula 2005).
These models transform a massive disk into a bar via dynamical processes alone.  The evolution
can be simulated via N-body models.  One new twist has been added by Saha et al. (2011), in which the formation
of a rapidly rotating bar spins up an existing ``classical'' bulge to the point where cylindrical rotation (normally only seen in bars) occur in the classical bulge.

It is noteworthy that the classical bulge formation models invoke mergers and star formation,
while the N-body models are purely dynamical, with the exception of the Combes (2009) model
in which a bar can be resurrected by the inflow of gas to form a massive disk.  However, one
key element missing from those models is ``gastrophysics''.   For example, while we know that
the bulge is a bar, we also know that it has an abundance gradient - a feature that cannot
occur via dynamical processes alone.   The physical processes that are likely to be responsible
for the gradient are supernovae and the outflow of metal enriched winds, along with dissipation
of the gas from which these generations of stars are formed.  N-body models do not presently
include these baryonic processes.

\subsection{ The Observational Landscape}

The bulge region has been long known to have a barred potential (e.g. Blitz \& Spergel 1991;
Burton \& Liszt 1993) and an asymmetric structure in the infrared, suggestive of a bar with
a long axis at an angle of $20-40^\circ$ from the line of sight between the Sun and the
Galactic Center.  The bulge is dominated by an old ($\sim$10 Gyr) population (Terndrup 1988; Ortolani et al. 1995; Kuijken
\& Rich 2002; Zoccali et a. 2003; Clarkson et al. 2008,
2011) with any young population in the bulge being $<3.4\%$ mass, by fraction, and likely less
(Figure 3).   The small population of stars securely identified as lying brighter than the globular
cluster-age turnoff are likely blue stragglers; Clarkson et al.  (2011) shows that least half are
true blue stragglers, with examples of clear W UMa variability.   The metallicity distribution was
first defined at low resolution by Rich (1988) and via high resolution spectroscopy  McWilliam
\& Rich (1994).  Rich (1990) fit the abundance distribution to the Simple Model of chemical evolution; while this is almost certainly an oversimplification, the presumption of an early almost instantaneous burst of star formation being responsible for the enrichment remains as the guiding principle for the bulge's chemical evolution, to the present day.  Recent studies confirm these earlier ones, finding a slightly subsolar mean
abundance, and extending to $\approx +0.5$ dex at the metal rich end.  In the first instance,
when Baade discovered large numbers of RR Lyrae stars in the bulge, it was considered to be
metal poor and globular cluster-like.  The perception then swung to metal rich.  In reality,
one may describe its abundance distribution as being like the disk, but more broad, extended
both to lower and higher metallicities.  As McWilliam \& Rich (1994) first pointed out, and
is confirmed by many other studies (e.g. Fulbright, McWilliam, \& Rich 2006, 2007), the bulge
stars are generally enhanced in alpha elements; Ballero et al. (2007) argue that this is
consistent with early, rapid, enrichment.  As pointed out for bulge stars in 
Baade's Window by McWilliam, Fulbright, \& Rich
(2010), the [La/Eu] ratio is close to the ratio consistent with the r-process, i.e enrichment via SNe rather
than the envelopes of intermediate-mass stars (Figure 4)

There is also a vertical abundance gradient outside of 500 pc, although there appears to be no gradient
within 500 pc (Rich et al. 2007).   Early work by Terndrup (1988) and
Tyson \& Rich (1991) found a vertical abundance gradient, and the abundances of red clump stars
(Zoccali et al. 2008) definitively show a minor axis vertical abundance gradient from $-4^\circ$ to $-12^\circ$, with a
new field ( $-8^\circ$; Johnson et al. 2011; Figure 7) consistent with the trend.  Abundance gradients are seen in edge-on spirals like NGC 4565 (Proctor et al. 2000), which is
considered to be a peanut shaped bulge.   

Surprisingly, prior to the BRAVA survey, only small samples of 
kinematic probes had been surveyed at optical wavelengths, with the majority of bulge dynamical probes
being the OH/IR and SiO masers, very late-type stars observed at radio wavelengths.  The new generation of surveys
has dramatically altered that situation.

The overall observational picture carries little evidence for any extended star formation
history (e.g. no known thermally pulsing carbon stars) and the alpha enhancement, supports a
$<1$ Gyr burst of star formation during which most of the chemical enrichment of the bulge
is thought to have occurred.  Although the secular evolution that is thought to produce a
massive bar is considered to be more extended in time, that process could also conceivably have
occurred within 1 Gyr, or the roughly 10-20 dynamical times for the inner Galaxy.  

One recent development has been observational claims for an ``X-shaped'' bulge.  McWilliam
\& Zoccali (2010) found a doubling of the red clump in a number of bulge fields.  
de Propris et al. (2011; Figure 5) find no quantitative differences in metallicity or kinematics,
across the two clumps.   If the doubled clumps have their origin in a purely dynamical process that
caused the bar to buckle and the ``X-structure'' to appear, then one might suspect it took place after the bulge
stars had already formed, and consequently one would not expect to see differences in metallicity between the
two clumps.  However, given the spatial separation, one rather expects to see measurable kinematic differences.
When larger samples are accumulated, the member population of
the X-structure should exhibit measurable differences in dynamics, when proper motion data
is combined with radial velocities.

\begin{figure*}
\centering
\includegraphics[width=0.90\columnwidth]{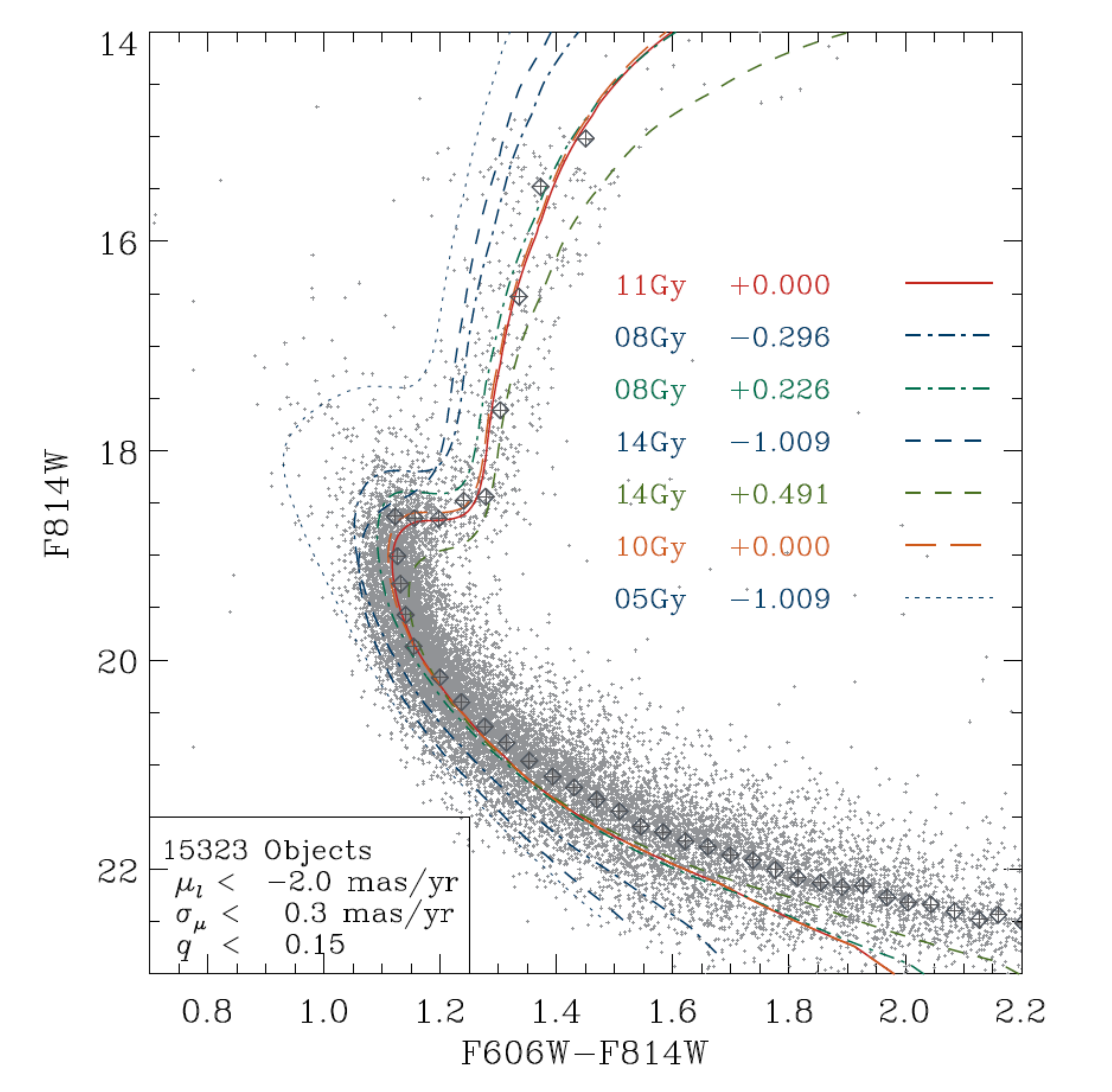}
\vskip0pt
\caption{Proper motion-selected bulge objects (to reject foreground disk stars)
from a deep HST/ACS dataset, using similar mean proper
motion criteria to Kuijken \& Rich (2002) but with a 6$\sigma$ detection requirement imposed
(Clarkson et al. 2008).
This CMD was divided into bins and the median computed (diamonds); below the MSTO
the uncertain binary fraction causes an artificial apparent age effect, so we focus on the
region above the MSTO for comparison. An alpha-enhanced, solar-metallicity isochrone at
11Gyr represents the median sequence well above the turn-off. Also shown are sequences
at metallicity [Fe/H]=$(-1.009,-0.226, +0.491)$ and ages (8, 10, 14) Gyr to bracket the Bulge
population above the MSTO. Also shown is a very young, very metal-poor population (dotted
line).  See Clarkson et al. (2008) for details.  The CMD admits very little possible
young/intermediate age population.}
\label{Clarkson CMD}
\end{figure*}

\begin{figure*}
\centering
\includegraphics[width=0.90\columnwidth]{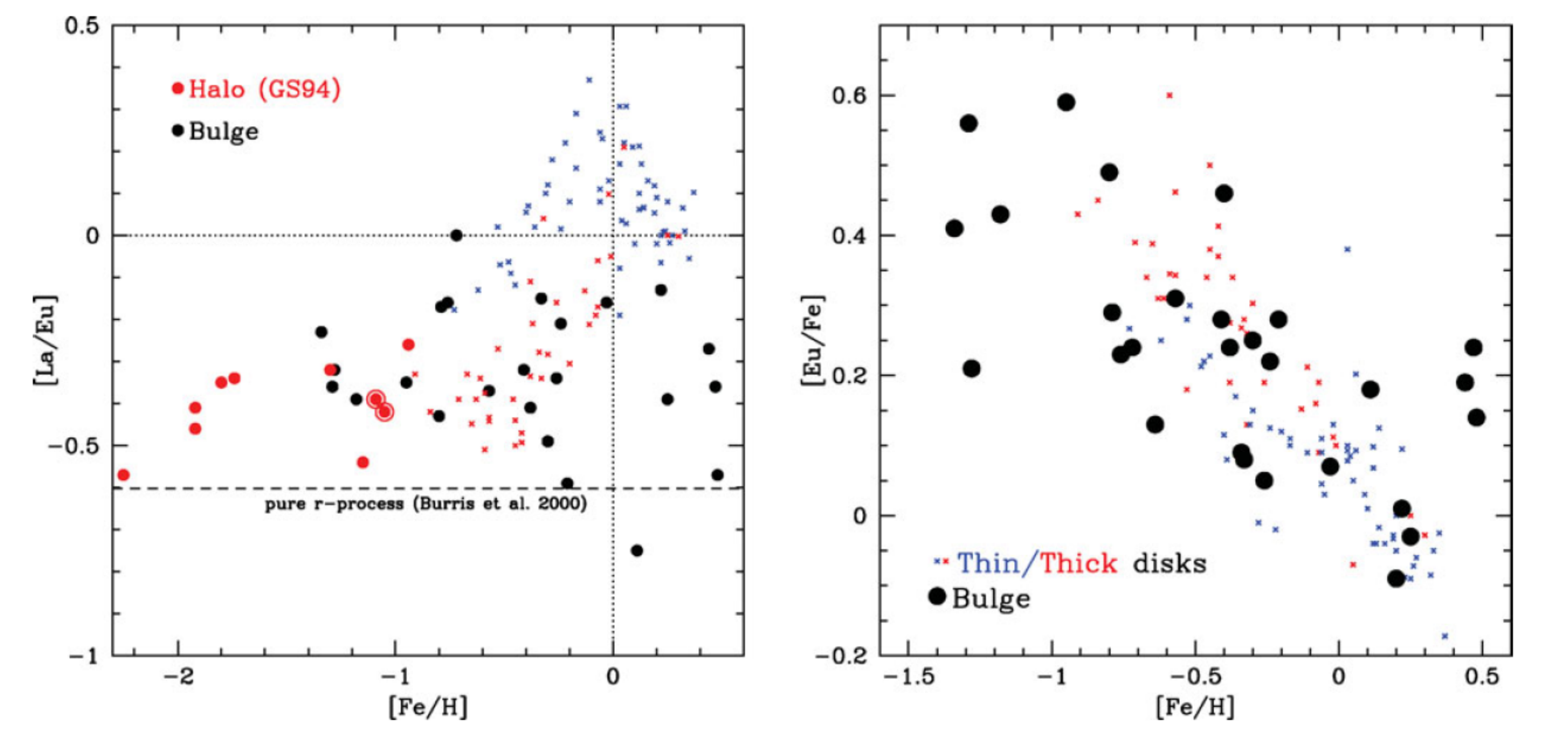}
\vskip0pt
\caption{ Heavy elements in Baade's Window $(b=-4^\circ)$, from McWilliam, Fulbright, \& Rich 2010.  (Left:) [La/Eu] in the bulge and disk; blue and red crosses are thin and thick disk respectively.  Notice that the bulge giants continue a halo-like trend to higher metallicity and even include a metal rich, strong r-process, star; other examples are found in the Plaut Field; Johnson et al. 2011.  (Right:)  [Eu/Fe] vs [Fe/H].  Although generally following a trend similar to the thin and thick disk, the 3 stars with elevated [Eu/Fe] at [Fe/H]=+0.5 are noteworthy.  If more such stars were found, it would suggest that the whole enrichment process, even to the most metal rich stars, was more rapid than even the thick disk.  }
\label{heavy elements}
\end{figure*}

\begin{figure*}
\centering
\includegraphics[width=0.90\columnwidth]{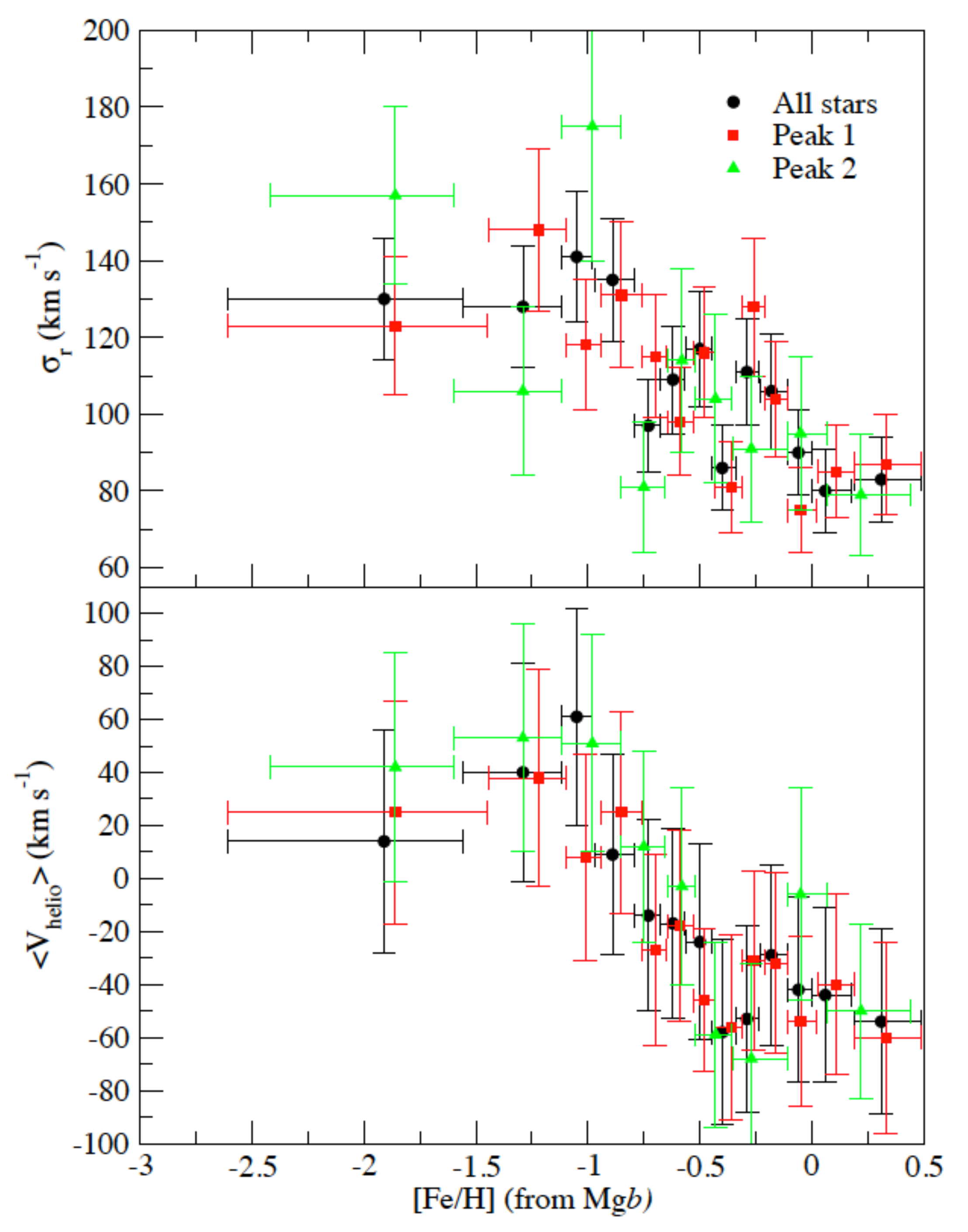}
\vskip0pt
\caption{ Spectroscopy of stars in the Plaut field, which exhibits a double red clump (McWilliam \& Zoccali 2010).     No difference is observed in kinematics or abundance for stars in the two clump populations in the bulge Plaut Field
$(l,b)=0^\circ, -8^\circ$ at 1 kpc (de Propris et al. 2011).
The brighter peak is plotted with red points.  If the X-shape is part of the general population bar dynamics (Athanassoula 2005) one would not necessarily expect to see any composition or age difference.  Likely, proper motion data will be needed to spot
the kinematic differences in these populations.  }
\label{de propris plot}
\end{figure*}


\section{ Major Results of the BRAVA Survey}

Because BRAVA surveyed the bulge in a grid spanning both latitude and longitude (Figure
 1) it has become possible to investigate the rotation field perpendicular to the plane.
  The BRAVA survey shows that the bulge departs from pure ``solid body'' rotation (Howard
 et al. 2008) and has cylindrical rotation (Howard et al. 2009; Rich et al. 2008, 2009),
 a characteristic of pseudobulges (Kormendy \& Kennicutt 2004).  The rotation field
 for the $-4^\circ$ and $-8^\circ$ slices is the same.  The survey has now covered
 most of the Southern half of the bulge, and none of the fields shows clear evidence
 for cold streams.   The most significant result (Figure 6) indicates that the fraction
 of the bulge mass in a ``classical'' non-barred configuration must be $<8\%$ (Shen et
 al. 2010).  Analysis of new data by Kunder et al. (2011) confirms the cylindrical
 rotation, finding it for the $-6^\circ$ field as well.   An additional result from
 the BRAVA study places the Milky Way in the Binney plot (Figure 7) and it falls next
 to the well known ``peanut shaped'' bulge galaxy, NGC 4565, lying slightly above the
 oblate rotator model line.   It is noteworthy that the posited ``X-shape'' feature in
 the bulge (McWilliam \& Zoccali 2010) emerged early on in N-body simulations (Combes \& Sanders 1981) and is modeled in detail in Athanassoula (2005).

We find no ``high velocity'' stars in the BRAVA sample.  One field with a promising
 $2\sigma$ cold stream was followed up with additional spectroscopy.  The result proved
 the validity of the Central Limit Theorem, as the possible ``stream'' disappeared when
 the sample size quadrupled (Howard et al. 2008).

The BRAVA database will be used to constrain a new self-consistent dynamical model
 along the lines of the one developed by Zhao (1996).  At present, the N-body bar model of Shen et al. (2010); Figure 6 is an excellent fit to the radial velocity dataset over the while of the bulge.  The Besancon starcount/dynamical
 model of the Milky Way is also
 being adapted to model the BRAVA data (Robin et al. 2012 in prep.).
 
The BRAVA primary kinematic sample is ill-suited for abundance studies because the
 stars are too red for application of the Ca triplet abundance indicator.  The 8430
 TiO band veils the first two Ca infrared triplet lines and renders any Ca triplet
 index measurement useless.     A. Koch has been attempting to use the TiO bands as
 a proxy abundance indicator, and has had some modest success in confirming the abundance
 gradient (Kunder et al. 2011) but there is significantly more investigation required
 before quantitative abundance information can be derived from this feature.
 
 The complete dataset
 for the BRAVA survey is ultimately going to be maintained on the IRSA archive, as
 well as {\tt http://brava.astro.ucla.edu/}; Kunder et al. (2011) describes the final low resolution
 BRAVA database.

\begin{figure*}
\centering
\includegraphics[width=0.90\columnwidth]{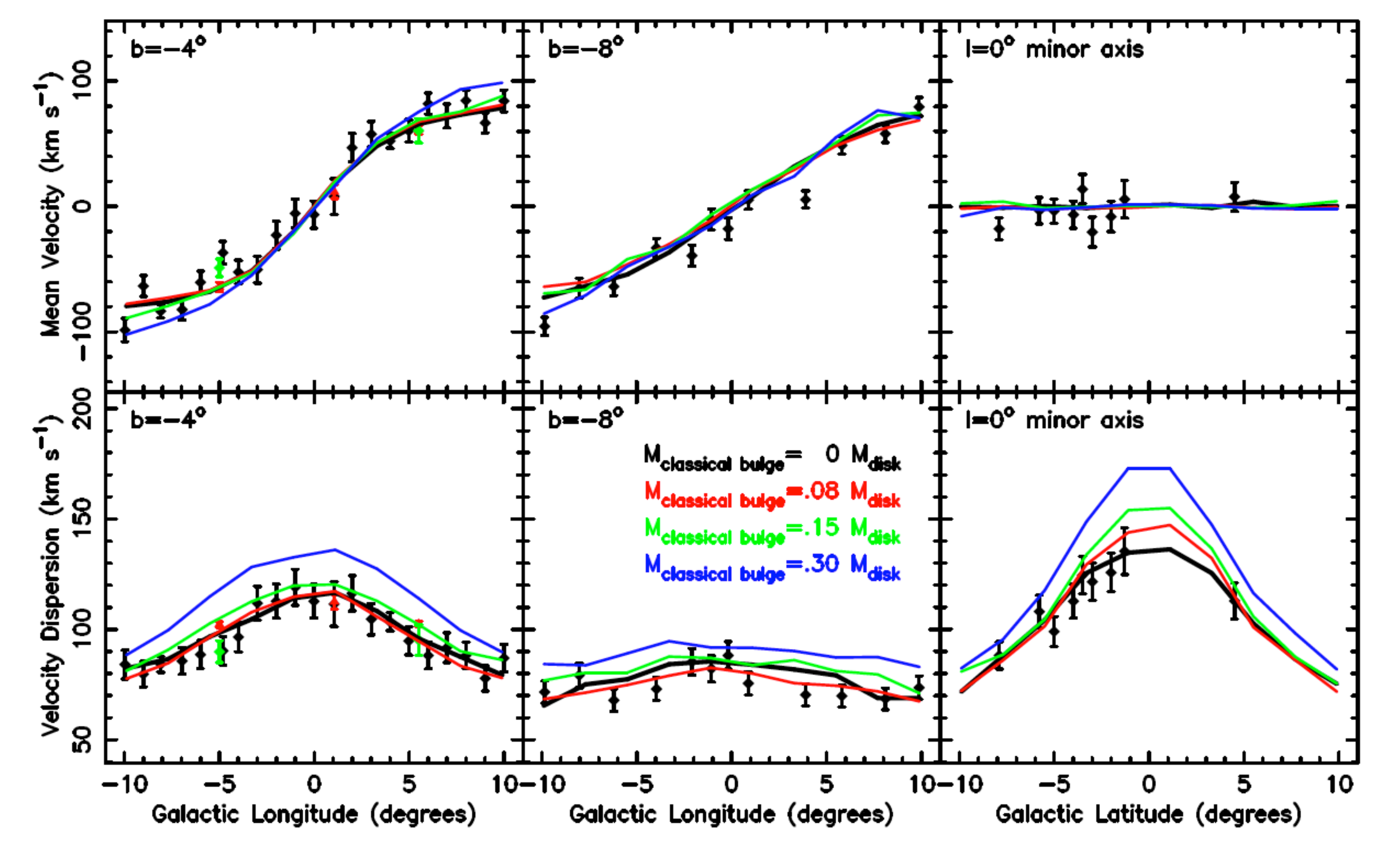}
\vskip0pt
\caption{Illustrates the fit of the N-body buckled bar model of Shen et al. (2010)
 to the BRAVA dataset, varying the fraction of mass in a hypothetical pre-existing
 classical bulge.  The heavy black lines are no classical bulge, while red, green,
 and blue lines indicate addition of a classical bulge with 8\%, 15\%, and 30\%  of
 the disk mass.  It is clear that no significant classical bulge is permitted by the
 dataset, especially in fitting the minor axis profile (Shen et al. 2010).
 }
\label{Shen}
\end{figure*}

\begin{figure*}
\centering
\includegraphics[width=0.90\columnwidth]{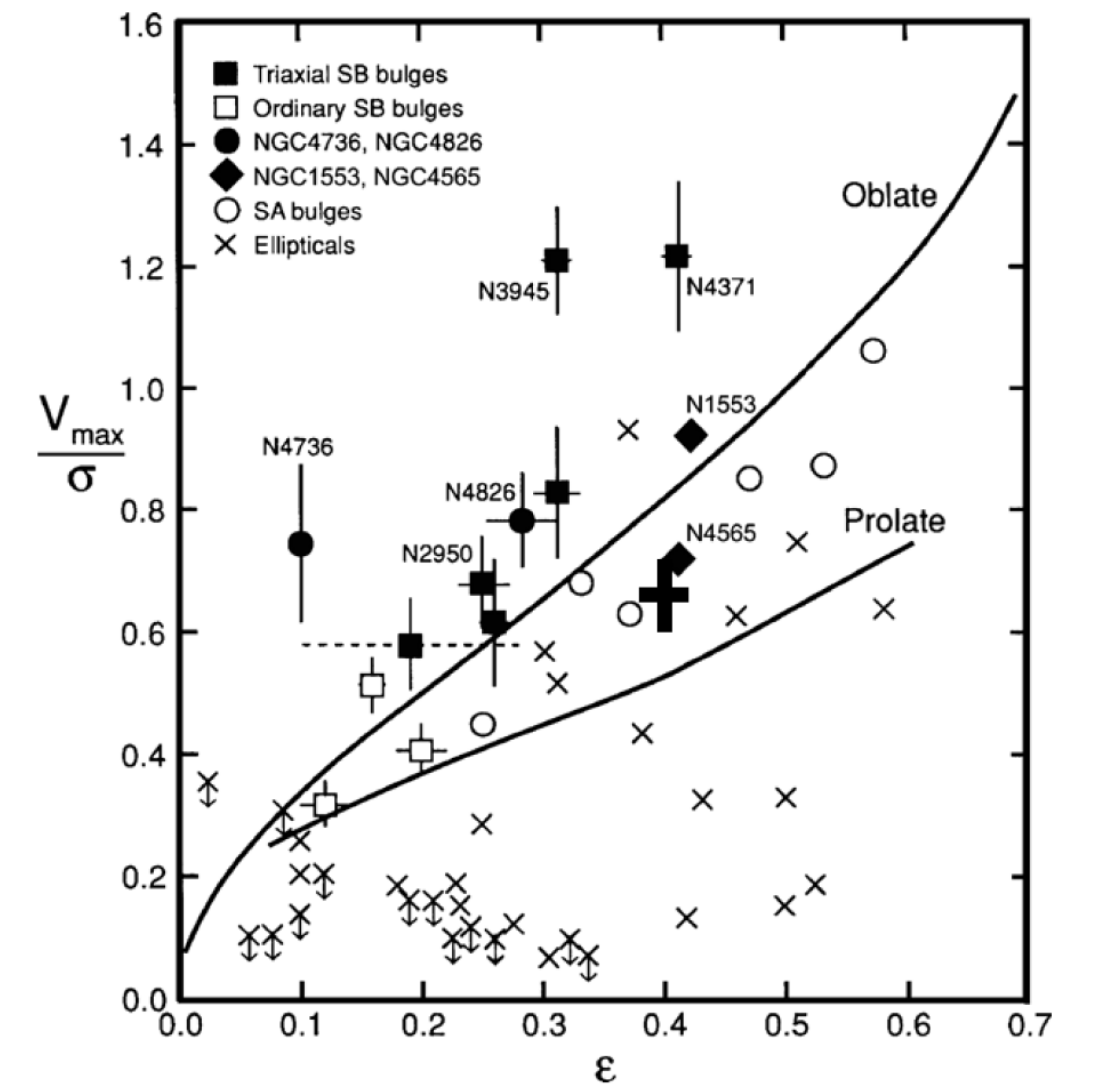}
\vskip0pt
\caption{ The ``Binney'' plot of $V_{max}/\sigma$ vs. eccentricity, from Kormendy \& Kennicutt (2004)
with the Galactic bulge BRAVA results indicated with a solid black cross.  The BRAVA result falls
below the oblate rotator line, nominally less rotationally
supported than the pseudobulges (filled symbols) but similar to classical bulges (open
symbols). NGC 4565 is an edge-on boxy peanut bulge, similar to the Milky Way; it has both
cylindrical rotation and an abundance gradient.}
\label{Binney}
\end{figure*}

\subsection{A new study of the Bulge Composition in the Plaut Field}

Although not strictly part of the BRAVA survey, we obtained multiobject high resolution
 echelle spectroscopy of clump giants in the Plaut bulge field $(l,b)=0^\circ, -8^\circ$
 beginning in 2007, using the hydra spectrograph at CTIO (Johnson et al. 2011).    Note
 that these stars are 2-3 mag fainter than the BRAVA primary sample, and are being
 observed at roughly 6 times as high spectral resolution.  Typical exposure times were
 8 hours for this sample, as opposed to 1 hr for the BRAVA M giant survey.   We have
 nearly completed a program to gather similar data in other bulge fields along the
 minor and major axes.  

The new study shows that the suspected abundance gradient found in the outer bulge
 by Zoccali et al. continues smoothly from $-6^\circ$ to $-8^\circ$.  This is an important
 finding (Figure 8) because Zoccali's outermost field, at $-12^\circ$, is arguably quite
 distant from the nucleus and may have a substantial inner halo representation.  In
 demonstrating that both the $-6^\circ$ and $-8^\circ$ fields follow the general rotation
 field, BRAVA shows that at 1 kpc from the nucleus, the population remains bulge-dominated. 

A further result is that the the the alpha elements remain elevated 1000 pc from the
 plane, any gradient in iron notwithstanding.  So the conditions giving rise to the
 high alpha abundances occurred over the full extent of the bulge; presumably the bulge
 formed rapidly, and violently, over that whole volume. Still, the existence of the
 gradient is consistent with dissipation and possibly winds playing a role in the chemical
 enrichment history.  Note that in the Simple model of chemical evolution, winds cause
 a decline in the mean abundance while preserving the shape (Hartwick 1976).  In fact,
 the shapes of the abundance distributions look remarkably similar.   This is the first
 clear demonstration that the alpha enhancement occurs over the entire volume of the bulge/bar,
 out to a 1 kpc radius.

In the course of the study, we also analyzed a population of red clump stars that appear
 to be much closer than the bulge, about 2 kpc distant.  Surprisingly, these stars
 also show elevated alpha abundances and have disk-like metallicity (reaching +0.4
 dex) and are kinematically disk like as well.  Note that this is {\it unrelated} to
 the doubled red clump populations, as these stars are both brighter and bluer than
 those, appearing to fall closer and less reddened than any bulge population.  It will
 be important to study this population more in the future.

Our team is just concluding analysis of the heavy elements.  We are confirming the
 high Eu abundance and r-process-like [La/Eu] ratios found by McWilliam, Fulbright,
 \& Rich (2010) in Baade's Window.  We are finding hints of new phenomena, including
 large scatter in the s-process, perhaps indicative of a late population of AGB stars
 in the metal poor halo.   We believe that the heavy elements will offer us the scalpel
 needed to explore the 1 Gyr time frame in which the bulge enrichment took place.

\begin{figure*}
\centering
\includegraphics[width=0.90\columnwidth]{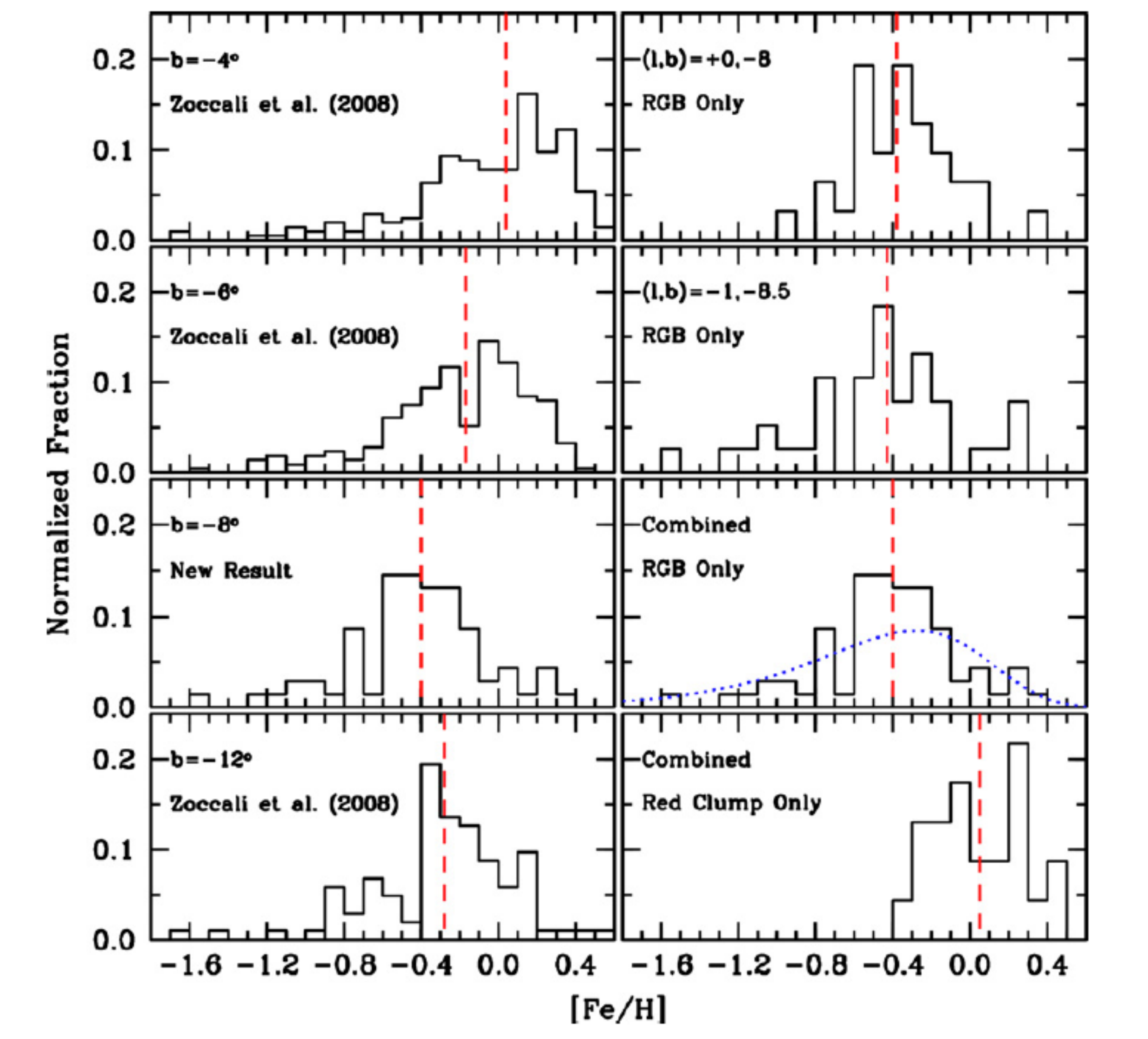}
\vskip0pt
\caption{ Metallicity distribution functions (high resolution) of bulge fields in 0.1 dex bins
(Johnson et al. 2011).  The same study shows that the bulge population remains alpha-enhanced at $b=-8^\circ$.
For all panels, the dashed red line designates the median [Fe/H]
value. The left panels compare the metallicity distribution functions of Zoccali et al. (2008)
to our combined spectroscopic data at $b=-8^\circ$ and $ b =-8.5^\circ$. The right
panels show the spectroscopic metallicity distribution functions for various subsamples in
the Johnson fields, including that of a foreground red clump population (lower right plot).
The dotted blue line shows the result of a one-zone, simple model calculation with
a yield of z = 0.0105. Area under the model curve has been scaled to equal the area under
the data histogram.}
\label{Johnson}
\end{figure*}

\begin{figure*}
\centering
\includegraphics[width=0.90\columnwidth]{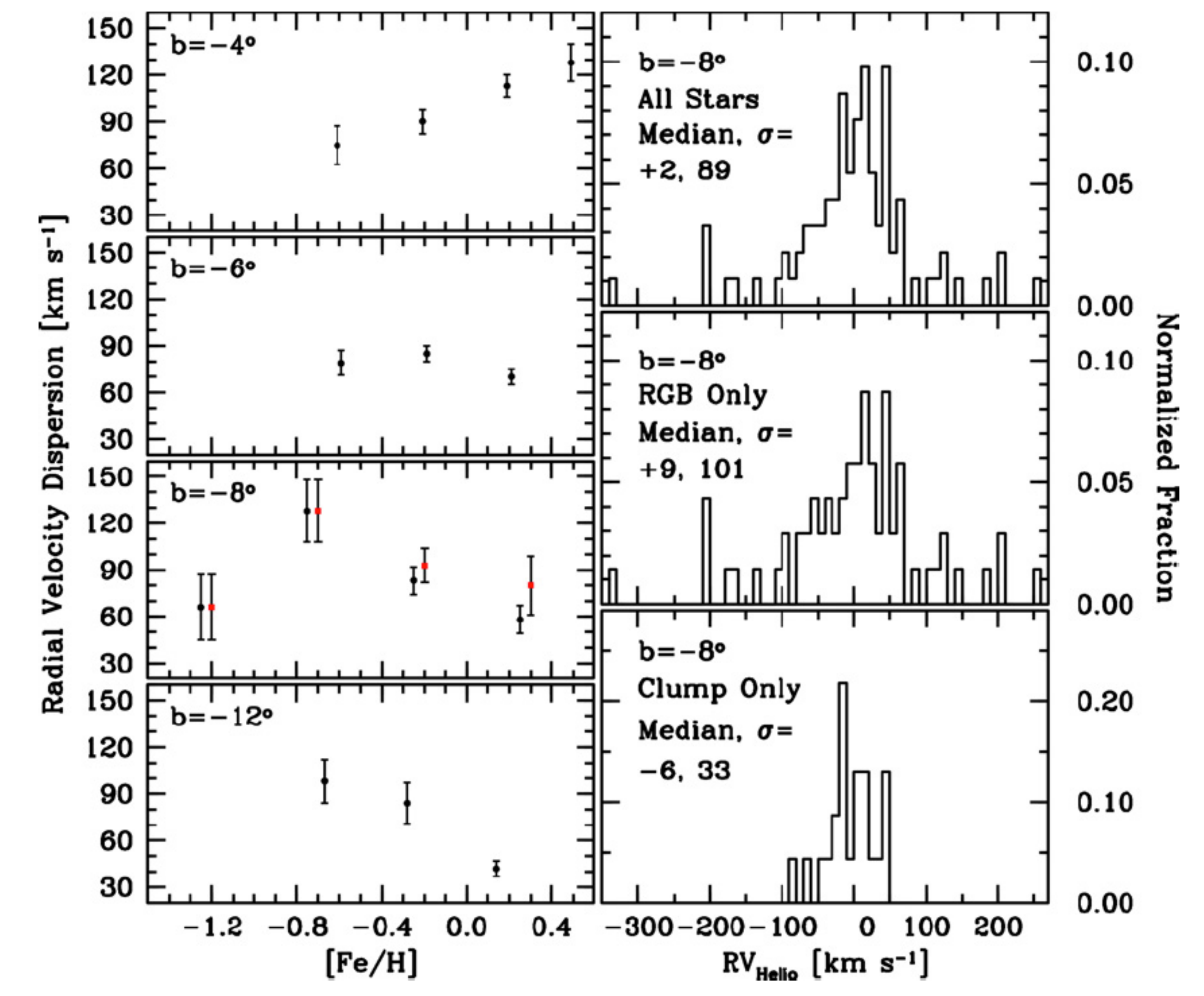}
\vskip0pt
\caption{The bulge is exhibiting a counter-intuitive trend of velocity dispersion with metallicity
in Baade's Window; this is a troubling example of population complexity.  It is not clear how to
explain the evolving trend with Galactic latitude; this figure from Johnson et al. 2011.  Left
panels show the measured radial velocity (RV) dispersion as a function of metallicity along the bulge
minor axis at b = $-4^\circ$, $-6^\circ$, $-8^\circ$ , and $-12^\circ$. The $b=-8^\circ$ data are
from Johnson et al. (2011; 0.5 dex [Fe/H] bins), and the other fields were taken from Babusiaux
et al. (2010; 0.4 dex bins). In the $b=-8^\circ$ panel, the filled black circles represent the RV
dispersion when the full sample is taken into account, and the filled red squares represent the RV
dispersion when only the RGB stars are used. The histograms in the right panels illustrate the RV
distribution at $b=-8^\circ$ with the full sample (top panel), the RGB stars only (middle panel), 
and the foreground red clump stars only (bottom panel) in 10 $km sec^{-1}$ bins; see Johnson
et al. 2011 for details.}
\label{Johnson-velocity}
\end{figure*}

\section{Terzan 5:  Curiosity or Clue?}

In the course of a routine verification program with a prototype multiconjugate adaptive
 optics imager at VLT, Ferraro et al. (2009) discovered that the old, metal rich, globular
 cluster Terzan 5 has a horizontal branch that is clearly bimodal in luminosity.    Terzan
 5 is remarkable, in that it is $9.5\times 10^5 L_\odot$  (Lanzoni et al. 2010) and
 boasts 34 msec pulsars - roughly 25\% of the known total.  Initial studies shed some
 light on the bimodal HB, a phenomenon unprecedented in any other globular cluster.
  The brighter HB stars were found to be both more centrally concentrated and more
 metal rich (the latter, based on spectroscopy using nirspec at Keck).  The metallicity
 difference was significant: the metal rich subgroup was found to have [Fe/H] $\approx
 +0.3$, 0.5 dex above the metal poor subgroup.  Stellar evolution models require that
 either the metal rich subgroup be 6 Gyr or helium enhanced.

When the sample size was increased, an even more remarkable feature emerged (Origlia
 et al. 2011).  Overall, the trend of $\rm [\alpha/Fe]$ vs [Fe/H] appears bulge-like,
 but there is striking bimodality in both the iron and alpha abundances (Figure 11).
  Neither subgroup displays the classic Na-O anti-correlation that is characteristic of globular clusters. 

We cannot explain this phenomenon.  We can be certain that it is not the merger of
 two globular clusters.  The velocities of the two subgroups are identical, as are
 the dispersions.  There is no known globular cluster at [Fe/H]=+0.3, so it would be
 unlikely that the only such cluster in the Milky Way merged with with another metal rich cluster.
A possibly more attractive hypothesis is that Terzan 5 is not a normal globular cluster
 at all, and is rather the remnant of some much larger ``building block'' system.  Since
 multiple populations are present, we can appeal to a proto-system that had significantly
 greater mass, able to contain the enriched gas produced in SNe.  Ferraro et al. propose
 that such special systems might have been the chemical building blocks of the bulge.
   We do not observe any tendency toward bimodality in bulge field population, although
 one could imagine similar systems with a range of properties.   In the case of Terzan
 5, it would be important to search for heavy element signatures that might be found
 in bulge field stars.  With improved instrumentation, one might also search for clues
 in other more subtle measurements, like isotope ratios of Mg.  

For the moment, Terzan 5 remains a fascinating system, and it is difficult to evaluate
 fully how important the phenomenon is.  The age difference between these two populations
 is a critical measurement, yet it is also one that time allocation committees have
 steadfastly refused to support.  So the mystery remains very much unsolved, and may
 offer new clues to the bulge's formation.

\section{ Complications}

The observational picture of the ``bulge'' or bar is becoming increasingly complicated, and there are a number of contradictions that are difficult to resolve.   The M giant and PNe kinematics appear to be consistent with an N-body bar model.   Yet a number of studies argue for a much more complicated picture.   Beginning with Zhao, Spergel, \& Rich (1994), and then Soto et al. (2007), the combination of abundances, proper motions, and radial velocities finds stars with metallicity higher than $-0.5$ dex exhibit ``vertex deviation'' while the more metal poor stars do not.   That is, when 
proper motion data are included with radial velocities, stars supporting the bar have a strong correlation between radial velocity and proper motion parallel to the plane.  Babusiaux et al. (2010) confirm the vertex deviation of  Soto et al. (2007) with a similar sample size.  However, they find a very strong trend of velocity dispersion with metallicity in Baade's Window, with the metal rich population having the highest velocity dispersion-a trend not noted in any other population.   Their metallicity-velocity dispersion trend becomes flat in the $-6^\circ$ field and reverses at $b=-12^\circ$.    It is interesting that Johnson et al. (2011; Figure 9) finds that same trend in the $b=-8^\circ$ field.   In a dissipative formation picture, one would expect that the most metal rich population would show the most ordered kinematics (lowest sigma), having formed from the most evolved gas.  It is difficult to understand their trends in Baade's Window; Babusiaux et al. (2010) propose that the metal rich ``hot'' population is more concentrated to the plane.  One might appeal to the long flat bar (e.g. Benjamin et al. 2005) but that feature is only significant close to the plane.  It would be important to confirm this trend in other bulge fields and to understand why it is not present in the $-6^\circ$ field.
De Propris et al. (2011) also see a strong abundance-kinematics correlation in the sense that metal poor stars have both higher velocity dispersion and slower rotation than the metal rich stars, although this study was at $b=-8^\circ$, in a region where Babusiaux et al. did not see their metallicity-velocity dispersion correlation.

I have alluded to another flavor of complexity, and that is in spatial geometry.  The doubling of the red clump and apparent ``X-shaped'' bulge, which has now been confirmed over the entire bulge (Saito et al. 2011).  Examining the doubled red clump plots in their paper, it is clear that the phenomenon is not merely a ``decoration'' but involves a significant fraction of the mass.  It is entirely possible that the X-shaped (or perhaps ``dog bone'' ) structure is part of the bar's expressed dynamics, and that we will not see strong composition differences in the different populations.  Indeed, Athanassoula (2005) argues that the X-shape is characteristic of an edge-on strong bar, and the simulations show that the X-shaped feature is extensive; her figures are worthy of inspection, and show how strong the X-shaped structure can be in these simulations.   If it turns out that the Milky Way has an X-shaped bulge, as is likely, then one may conclude that the observed dynamics in the BRAVA survey- especially the cylindrical rotation - is a ``genuine'' reflection of a bar, rather than a spun-up classical bulge (e.g. Saha et a. 2011).  As the photometric data improve, it will be vital to fit both the dynamics and the X-structure.  An important future goal will be to apply astrometry and precision spectroscopy to improve the placement of the red clump stars in space.  The proposed Japanese infrared counterpart to GAIA, called Jasmine, is capable of measuring proper motions for every red clump star in the bulge --an exercise that of course is possible only in our Galaxy.   Given the remarkable developments concerning the structure of the bulge, Jasmine should now be considered to be a critical mission; GAIA is an optical mission and will not be capable of studying the faint doubled clump population.

Yet another proposed form of complexity is the deconvolution of the metallicity distribution function into multiple populations.  Considering the sample of 26 microlensed dwarfs studied by many different groups, Bensby et al. (2011) find a bimodal abundance distribution, with Solar metallicity stars lacking.  The metal rich peak also includes stars with age (derived from a spectroscopic parallax) of $\sim 6 $ Gyr, fortuitously coincidental with the age of the younger population in Terzan 5.   The microlensing results could reflect small number statistics, or we might be observing a population mix if the sources are not exactly at the bar center (if due to microlensing geometry).  However, one might expect such strong metallicity bimodality to be observable in the color-magnitude diagram (note that Brown et al. 2010 does see hints of bimodality in the main sequence turnoff, but attributes it to the color-metallicity relation rather than a physical cause).    

Hill et al. (2011) present a novel deconvolution of the metallicity distribution of K giants (the original Zoccali (2008) sample) that approximately recovers the two peaks of Bensby's microlensing sample.  However, the {\it same} sub populations appear to correspond to a metal rich ``bar'' population that has a vertex deviation, and a metal poor ``bulge'' population that does not.   Indeed, the vertex deviation suddenly setting in at [Fe/H]$\sim -0.5$ appears to be consistent with this deconvolution, which is physically supported by the microlensing sample.  However, the strong rotation curve and cylindrical rotation at $b=-8^\circ$ would appear to contradict the presence of a significant metal poor classical bulge component at 1 kpc, where one expects it to dominate.  As the bulge structure is traced to low surface brightness, one notes boxy isophotes, not the predominance of an $r^{1/4}$ bulge.    This issue remains ripe for investigation, and it is completely possible that there are multiple structures all occupying the same volume.    A kind of ``warning'' comes from the studies of $\omega$ Cen, that strong population composition differences are not necessarily accompanied by spatial or kinematic differences.   And in our bulge, there may be multiple bars (as is seen in extragalactic cases), and these may not exhibit age or abundance differences.   There is also the preliminary work of the AAOMEGA bulge survey of Ness \& Freeman (2012 in preparation), that finds an even greater number of subcomponents in the metallicity distribution function.  The Hill et al. (2011) result is noteworthy because the peaks are accompanied by strong kinematic differences.

I have alluded earlier to the issue of the abundance gradient.  The vertical gradient is well established for $b<-4^{\circ}$ but appears to be absent from $-1^{\circ}  < b <-4^{\circ}$ (Rich et al. 2007; Rich et al. 2012; Figure 10).  However, Brown et al. (2010) argue that the fraction of metal rich stars increases inward of Baade's Window.  Due to the high extinction, one must rely on M giants, which give a result at variance with optical studies; the mean [Fe/H] is -0.1, with $\sigma=0.1$.  While one may appeal to mass loss as the culprit for the unusual M giant abundance distribution, strong depletion of the red giant branch would require that there be a large population of extreme HB or UV-bright stars; no such extraordinarily large population seems to be present (Terndrup et al. 2004) though the subject is worthy of additional study.  It is difficult to construct an N-body bar scenario with any abundance gradient, much less two regions with different gradients.

Ultimately, in considering ascending levels of complexity in the bulge, we will need large samples with distance and composition information.  Although the most recent chemical evolution model favors a single, rapid, chemical evolution/formation event (Cescutti \& Matteucci 2011), and the BRAVA survey is consistent with a rapidly rotating bar, the fact of the vertical abundance gradient and cylindrical rotation are strong indications that the population character and formation history is incredibly complex.  Not to mention the X-shaped structure.   Thus, the metallicity distribution may well be fit by multiple components, and the ``simple'' model of chemical evolution is probably not going to be in serious consideration, going forward.

\begin{figure*}
\centering
\includegraphics[width=12cm]{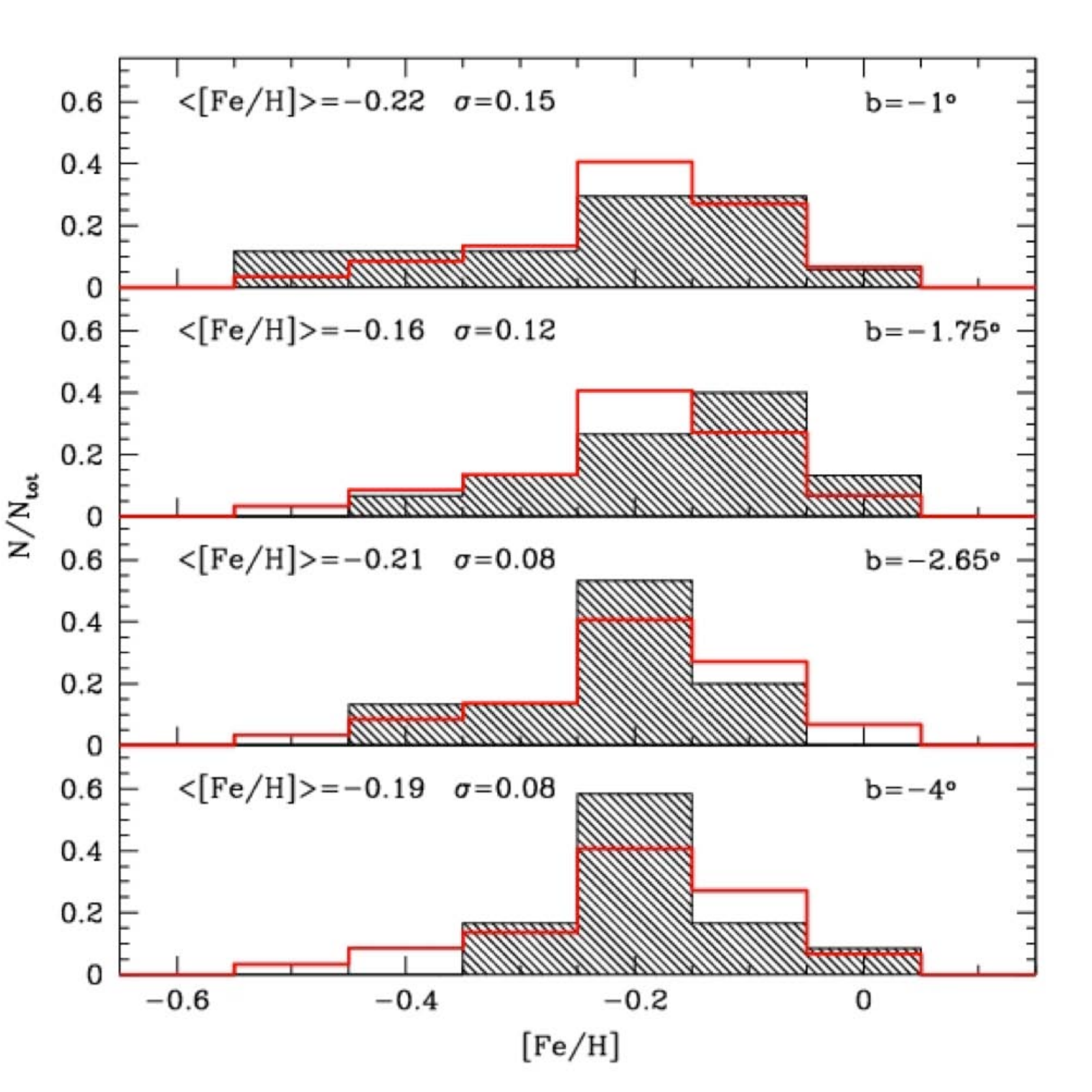}
\vskip0pt
\caption{ Composition of bulge M giants along the minor axis for $b<-4^\circ$ do not show an
abundance gradient (Rich, Origlia, \& Valenti 2012 in prep).  The red histogram is the co-added
distribution function.  The mean [Fe/H]=$-$0.1, and $\sigma=0.11$; both values are smaller than
is seen for K giant samples in Baade's Window.  Abundances are derived from R=25,000 nirspec 
spectra in the infrared H-band at 1.6$\mu$m (See Rich, Origlia, \& Valenti 2007).  The stars are
alpha enhanced, and there is no gradient in alpha-enhancement.   It is not immediately obvious
why there is an abundance gradient outside of Baade's Window, and it is possible that those
stars evolving to M giants suffer an abundance bias; this problem requires more investigation. }
\label{inner-gradient}
\end{figure*}

\section{ Conclusions}

The BRAVA survey has shown that the kinematics of 10,000 late M giants are consistent
 with the bulge being dominated by a bar population, with less than 8\% of the mass
 in a ``classical'' non-barred, bulge.  For the first time, we have seen cylindrical
 rotation in the bulge.  We note similarities to NGC 4565, which has a peanut-shaped
 bulge, cylindrical rotation, and an establish abundance gradient in its bulge, perpendicular
 to the plane (Proctor et al. 2000).   Considering the strong signatures that the bulge is in 
 fact a bar, we suspect that the X-shaped structure will not exhibit strong departures in composition
 or kinematics because it is an expression of the dynamics of the general population.  By the
 same token, the X-structure poses difficulties for models involving spun up classical bulges, which
 likely should not have such a structure. 
 
We also observe the signature of rapid bulge formation, alpha enhancement, over the
 entirety of the bulge's extent- or at least to $b=-8^\circ$ or 1 kpc.   One question
 is whether there is an abundance gradient interior to Baade's Window.  Infrared studies
 (Rich et al. 2005; Origlia \& Rich 2007) find no abundance gradient in the inner bulge.
  It will be vital to verify these findings, as they may give us important clues as
 to how the bulge formed.
 
 Many significant problems remain, especially how it is possible that one observes a strong abundance
 gradient in a population that exhibits the dynamical characteristics one expects of an N-body bar that has
 evolved only through dynamical processes.   The bulge continues to be of interest because it is the only such population where one has hope of measuring proper motions, detailed compositions, and even the spatial
 distribution of the stars.  Our bulge/bar offers the best hope of getting deep insight into the formation process
 of such systems.

New surveys of the bulge are underway.  APOGEE will examine the Northern bulge, and
 will study M giants in the infrared; it has begun in 2011.  Another survey at AAO (Ness \& Freeman 2012)
 targets the fainter clump giants.  These surveys will soon superseded BRAVA, adding yet more to our 
 understanding of the bulge.

\begin{figure*}
\centering
\includegraphics[width=0.95\columnwidth]{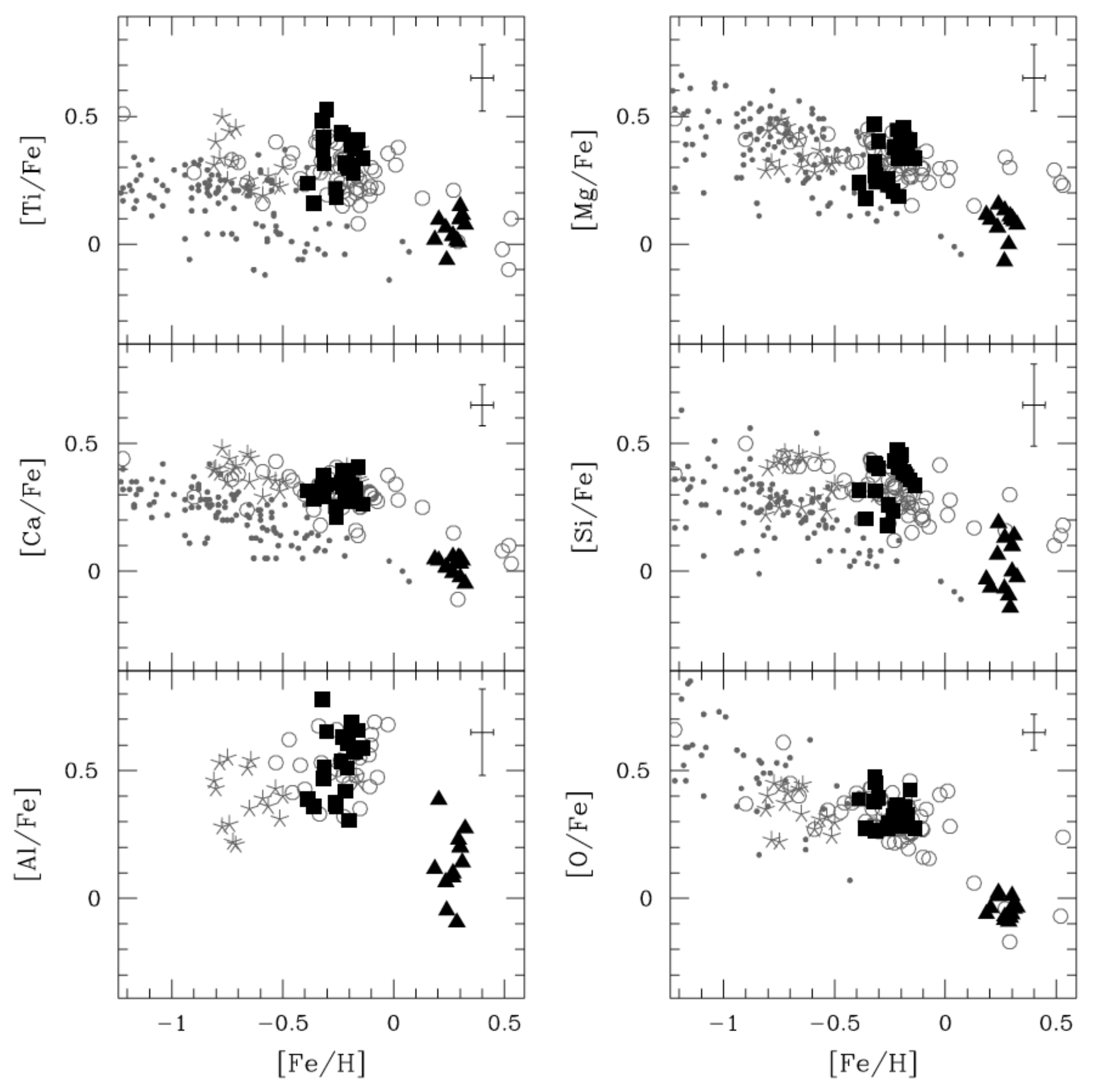}
\vskip0pt
\caption{ Remarkable bimodality of red giants in the bulge globular cluster Terzan
 5 (Origlia et al. 2011).  The solid symbols are red giants in Terzan 5, which exhibit
 bimodality in iron and all alpha elements- behavior unprecedented for any stellar
 system.  Other symbols illustrate various bulge field samples that show a more or
 less uniform distribution across [Fe/H].   Terzan 5 evidently experienced multiple
 generations of enrichment, but the metal rich subpopulation must also be either younger
 or more helium enhanced, because its horizontal branch population is brighter.  See
 Origlia et al.(2011) for details. }
\label{Origlia}
\end{figure*}

How much progress in the understanding of the bulge has been made since the beginning
 of George Preston's career, roughly at mid-20th Century?  It is remarkable to consider
 that Baade's (1958) discovery of RR Lyrae stars in the Galactic bulge from Mt. Wilson
 was the discovery that defined the bulge as a stellar population (old, like
 the globular clusters).  It was also the beginning of George Preston's career (and
 incidentally, the seminal 1957 Vatican symposium on Stellar Populations was held in
 the year of the author's birth).  The presence of {\it both} RR Lyrae stars and M
 giants in the bulge was noted at the Vatican Symposium, and the bulge was actually
 not classified with the oldest population II (globular clusters), but rather with
 the old disk. Well, as they say, what is old is new again.   Bensby et al. (2010) have suggested that the bulge's abundance pattern is more consistent with the thick disk, and a goal of the abundance survey of BRAVA
 is to test this hypothesis out by examining the composition of thick disk stars at
 the distance of the bulge.   We can point at the present time to far greater detail
 than was imaginable at the start of George's career:  color-magnitude diagrams based
 on proper-motion separation with the Hubble Space Telescope, and composition measurements
 using infrared echelle spectrographs.  We should not forget the surveys of variable
 stars and microlensing in the bulge, carried out at Las Campanas and thanks in part to his
 encouragement and sponsorship.    As George has
 so beautifully demonstrated with his high spectral resolution studies of RR Lyrae
 stars across their light curves, the very best data on even so \``mundane\'' a topic as
 nearby RR Lyrae stars offers a level of beauty and complexity that pushes beyond the
 capability of theory to offer an explanation.  We can hope that the next decades will
 offer data of that quality, in the Galactic bulge.
 
 \section{Acknowledgements}
 
 The author acknowledges support from AST-0709479 from the National Science Foundation, and thanks the Aspen Center for Physics where part of this paper was written.


\end{document}